\documentclass[aps,preprint,nofootinbib]{revtex4}
\usepackage{graphics}
\usepackage{slashed}
\usepackage{amssymb}
\usepackage{lscape}
\usepackage{amsmath}
\usepackage{graphicx}
\graphicspath{ {images/} }
\begin{document}

\title{Quantum thermodynamics in a static de Sitter space-time and initial state of the universe}
\author{$^{2}$ Juan Ignacio Musmarra\footnote{jmusmarra@mdp.edu.ar}, $^{1,2}$ Mauricio Bellini.
\footnote{{\bf Corresponding author}: mbellini@mdp.edu.ar} }
\address{$^1$ Departamento de F\'isica, Facultad de Ciencias Exactas y
Naturales, Universidad Nacional de Mar del Plata, Funes 3350, C.P.
7600, Mar del Plata, Argentina.\\
$^2$ Instituto de Investigaciones F\'{\i}sicas de Mar del Plata (IFIMAR), \\
Consejo Nacional de Investigaciones Cient\'ificas y T\'ecnicas
(CONICET), Mar del Plata, Argentina.}

\begin{abstract}
Using Relativistic Quantum Geometry we study back-reaction effects of space-time inside the causal horizon of a static de Sitter metric, in order to make a quantum thermodynamical description of space-time. We found a finite number of discrete energy levels for a scalar field from a polynomial condition of the confluent hypergeometric functions expanded around $r=0$. As in the previous work, we obtain that the uncertainty principle is valid for each energy level on sub-horizon scales of space-time. We found that temperature and entropy are dependent on the number of sub-states on each energy's level and the Bekenstein-Hawking temperature of each energy level is recovered when the number of sub-states of a given level tends to infinity. We propose that the primordial state of the universe could be described by a de Sitter metric with Planck energy $E_p=m_p\,c^2$, and a B-H temperature: $T_{BH}=\left(\frac{\hbar\,c}{2\pi\,l_p\,K_B}\right)$.
\end{abstract}
\maketitle

\section{Introduction and Motivation}

A de Sitter space-time is the maximally symmetric vacuum solution of Einstein's field equations with a positive cosmological constant $\Lambda $, which corresponds to a positive vacuum energy density and negative pressure. In the cosmological context, it describes the exponential accelerated
expansion of the universe governed by the vacuum energy density. There is evidence that the very early universe had a period of rapid expansion, called inflation\cite{infl,infl1,infl2,bcms}, well approximated by de Sitter space-time. Our tiny present-day cosmological constant currently accounts for about $68 \, \%$ of the energy density of the universe, and this fraction is growing as the universe continues to expand.  This means that we are entering a second de Sitter phase.  The early, inflationary de Sitter phase had a large cosmological constant and correspondingly tiny radius of curvature.  The future dark energy de Sitter will have an energy set by today’s cosmological constant, and enormous radius of curvature close to today’s Hubble scale\cite{mab}.

In the standard relativistic description, matter (which is described by the matter Lagrangian: $\hat{\cal L}$, in the Einstein-Hilbert (EH) action), is responsible for the spatial curvature of space-time, which is represented in the Einstein's equations through $G_{\alpha\beta}$. However, this description only takes into account the expectation value in the physical system under consideration. A better description must consider the effects that quantum fluctuations produce in the background space-time due to the retro-reaction, due to the fields that we are considering in $\hat{\cal L}$. Such description must be non-pertubative, because these effects could be very important when we deal with strong fields, under extreme physical conditions. Because of this, back-reaction effects are very important in general relativity, and in particular, they are essential to make a correct and accurate description of the initial state of the universe, which is believed to be given by a de Sitter metric. In this work we shall make a semiclassical description of such back-reaction effects, without consider non-commutative aspects of space-time, or the origin of quantum spinor fields that originate these quantum effects\cite{og,ab}. However, our semiclassical treatment should be sufficient to describe correctly the quantum thermodynamics inside the horizon of a de Sitter metric. A 4D de Sitter space is an Einstein manifold since the Ricci tensor is proportional to the metric: $R_{\mu\nu} = {3 \over  \alpha^2} \, g_{\mu\nu}$. It describes a vacuum solution of the Einstein's equations with a cosmological constant given by $\Lambda = {3 \over  \alpha^2}$ and a scalar curvature $R=4 \Lambda=12/\alpha^2$, such that $\alpha$ is the cosmological horizon. Therefore, a de Sitter space-time describes an hyperbolic space for $r<\alpha$.

Many years ago, Bekenstein has argued that isolated stable thermodynamic systems in asymptotically flat space-times satisfies the universal entropy bound\cite{bekk}: $S \leq {2\pi \alpha E \over \hbar c}$, where $\alpha$ is the radius of an enclosed system with energy $E$.
In this work we shall use a recently introduced thermodynamic description of space-time\cite{mba} in the study a Schwarzschild black-hole, but now with the aim to explore the interior of a de Sitter space-time (i.e., in the range $r<\alpha$). We shall use the formalism of Relativistic Quantum Geometry (RQG) described in \cite{rb} and \cite{rb1}, which was revisited in Sect. \ref{2}. In Sect. \ref{3} we study back-reaction effects inside the causal horizon of a de Sitter metric, with the aim to explore a quantum thermodynamical description of energy, lengths, entropy and the temperature.
Finally, in Sect. \ref{4}, we develop some final comments and conclusions.

\section{Revisited back-reaction effects from boundary conditions in the variation of the EH action}\label{2}

It is known that in the event that a manifold has a boundary
$\partial{\cal{M}}$, the action should be supplemented by a
boundary term for the variational principle to be
well-defined\cite{1,2}. However, this is not the only manner to
study this problem. As was demonstrated in\cite{rb,rb1}, there
is another way to include the flux around a hypersurface that
encloses a physical source without the inclusion of another term
in the Einstein-Hilbert (EH) action
\begin{equation}
 S_{EH} = \frac{1}{2k} \int d^{4}x \, \sqrt{-g} \,\left[\frac{\hat{R}}{2\kappa} + \hat{{\cal L}}\right],
\label{eqn:EinsteinHilbert}
\end{equation}
by making a constraint on the first variation of the EH action
\begin{equation}\label{delta}
 \delta S_{EH}  = \int d^{4}x \sqrt{-\hat{g}} \left[ \delta g^{\alpha \beta} \left( \hat{R}_{\alpha \beta} - \frac{g_{\alpha \beta}}{2} \hat{R} + \kappa\, \hat{T}_{\alpha \beta} \right) + \hat{g}^{\alpha \beta} \delta R_{\alpha \beta} \right] =0,
\end{equation}
where $\kappa=8\pi G/c^4$, $\hat{T}_{\alpha\beta}=2\frac{\delta \hat{{\cal L}}}{\delta g^{\alpha\beta}}-\hat{g}_{\alpha\beta} \hat{{\cal L}}$ is the stress tensor that describes matter and $\hat{{\cal L}}$ is the Lagrangian density. The last term in (\ref{delta}) is very important because takes into account boundary conditions. When that quantity is zero, we obtain the well known Einstein's equations without cosmological constant. This element can be written as:
\begin{equation}\label{bor}
\hat{g}^{\alpha\beta} \delta R_{\alpha\beta} =
\left[\delta W^{\alpha}\right]_{|\alpha} - \left( \hat{g}^{\alpha\epsilon} \right)_{|\epsilon}  \,\delta\Gamma^{\beta}_{\alpha\beta} +
 ( \hat{g}^{\alpha\beta} )_{|\epsilon}  \,\delta\Gamma^{\epsilon}_{\alpha\beta},
\end{equation}
where $\hat{g}^{\alpha\beta} \delta R_{\alpha\beta} =\delta \Phi(x^{\alpha}) = {\Lambda}\, \hat{g}^{\alpha\beta} \delta
g_{\alpha\beta}$ is the flux of the 4-vector $\hat{\delta W}^{\alpha}= \hat{\delta\Gamma}^{\epsilon}_{\beta\epsilon} \hat{g}^{\beta\alpha}- \hat{\delta \Gamma}^{\alpha}_{\beta\gamma} \hat{g}^{\beta\gamma}$ that cross any $3D$ closed manifold defined on an arbitrary region of the background manifold, which is considered as Riemannian and is characterized by the Levi-Civita connections. As in a previous work\cite{mba} we must describe the variation of the connections with respect to the background manifold, which is a Riemannian one. We shall consider no-metricity on the extended manifold. To extend the Riemann manifold we shall consider the connections
\begin{equation}\label{ConexionWeyl}
\Gamma^{\alpha}_{\beta\gamma} = \left\{ \begin{array}{cc}  \alpha \, \\ \beta \, \gamma  \end{array} \right\} + \delta \Gamma^{\alpha}_{\beta\gamma} = \left\{ \begin{array}{cc}  \alpha \, \\ \beta \, \gamma  \end{array} \right\}+ \beta \,\sigma^{\alpha} g_{\beta\gamma}.
\end{equation}
The last term is a geometrical displacement $\delta \Gamma^{\alpha}_{\beta\gamma}=\beta \,\sigma^{\alpha}\,g_{\beta\gamma}$ with respect to the background (Riemannian) manifold, described with the Levi-Civita connections. The particular case $\beta=1/3$ guarantees the integrability of boundary terms in (\ref{bor}). Here, $\sigma(x^{\alpha})$ is a scalar field. In that follows we shall denote: $\sigma_{\alpha}\equiv \sigma_{,\alpha}$ as the ordinary partial derivative of $\sigma$ with respect to $x^{\alpha}$.

The flux that cross the 3D-gaussian hypersurface, $\delta\Phi$, is related to the cosmological constant and the variation of the scalar field: $\delta\sigma$:
\begin{equation}
\delta\Phi = - \frac{4}{3} \Lambda\,\delta \sigma.
\end{equation}
In order for calculate $\delta R_{\alpha \beta}$, we shall use the Palatini identity\cite{pal}
\begin{equation}
\delta{R}^{\alpha}_{\beta\gamma\alpha}=\delta{R}_{\beta\gamma}= \left(\delta\Gamma^{\alpha}_{\beta\alpha} \right)_{| \gamma} - \left(\delta\Gamma^{\alpha}_{\beta\gamma} \right)_{| \alpha}.
\end{equation}
A very important fact is that the fields $\delta \bar{W}^{\alpha}$ are invariant under gauge-transformations $\delta \bar{W}^{\alpha} = \delta W^{\alpha} - \nabla_{\alpha} \delta \Phi$, where $\delta \Phi$ satisfy $\Box \delta \Phi=0$. Due to this fact, it is possible to define the Einstein's tensor trasformation $\bar{G}_{\alpha \beta} = G_{\alpha \beta} - \Lambda g_{\alpha \beta}$, which preserves the EH action
\begin{equation}\label{DinamicaMod}
\bar{G}_{\alpha \beta} = - \kappa\, \hat{T}_{\alpha \beta}.
\end{equation}
The condition of integrability expresses that we can assign univocally a norm to any vector in any point, so that it must be required that $\hat{g}^{\alpha\beta} \delta R_{\alpha\beta}=\nabla_{\alpha} \delta W^{\alpha}$. Of course, this is a particular case of (\ref{bor}). In particular, the case $\hat{g}^{\alpha\beta} \delta R_{\alpha\beta}=0$, gives us the standard Einstein's equations: $ \hat{G}_{\alpha\beta}+\kappa\,\hat{T}_{\alpha\beta}=0$.

In this background must be fulfilled: $\Delta g_{\alpha \beta} = \hat{g}_{\alpha \beta ; \gamma} dx^{\gamma}=0$. However, on the extended manifold, we obtain
\begin{equation}\label{VariaciongWeyl}
\delta g_{\alpha \beta} = \hat{g}_{\alpha \beta | \gamma} dx^{\gamma} = - \frac{1}{3} (\sigma_{\beta} \hat{g}_{\alpha \gamma} + \sigma_{\alpha} \hat{g}_{\beta \gamma}) dx^{\gamma},
\end{equation}
where $\hat{g}_{\alpha \beta | \gamma}$ denotes the covariant derivative on the extended manifold, once $\hat{g}_{\alpha \beta ; \gamma}=0$.
Therefore, the variation of the Ricci tensor on the extended manifold will be
\begin{equation}\label{VariacionRicciWeyl}
\delta R_{\alpha \beta}  = \left( \delta \Gamma^{\epsilon}_{\alpha \epsilon} \right)_{|\beta} - ( \delta \Gamma^{\epsilon}_{\alpha \beta} )_{|\epsilon} \\
 = \frac{1}{3} \left[ \nabla_{\beta} \sigma_{\alpha} + \frac{1}{3} \left( \sigma_{\alpha} \sigma_{\beta} + \sigma_{\beta} \sigma_{\alpha} \right) - \hat{g}_{\alpha \beta} \left( \nabla_{\epsilon} \sigma^{\epsilon} + \frac{2}{3} \sigma_{\nu} \sigma^{\nu} \right) \right],
\end{equation}
such that the variation of the scalar curvature is: $\delta R = \nabla_{\mu} \delta W^{\mu} = \nabla_{\mu} \sigma^{\mu} + \sigma_{\mu} \sigma^{\mu}$. The cosmological constant $\Lambda$ is an invariant on the background manifold, but not on the extended one: $\Lambda(\sigma,\sigma_{\alpha})=-\frac{1}{4} \left( \sigma_{\alpha} \sigma^{\alpha} + \Box \sigma \right)$, on which behaves as a functional. By defying the action
\begin{equation}\label{AccionLambda}
\mathcal{W} = \int d^{4}x \sqrt{-\hat{g}} \,\Lambda(\sigma, \sigma_{\alpha}).
\end{equation}
If we requiere that $\delta \mathcal{W} = 0$ we obtain that $\sigma$ is a free scalar field on the extended manifold: $\Box \sigma = 0$. The scalar field $\sigma$ describes the back reaction effects which leaves invariant the action:
\begin{equation}
{\cal S}_{EH} = \int d^4 x\, \sqrt{-\hat{g}}\, \left[\frac{\hat{R}}{2\kappa} + \hat{{\cal L}}\right] = \int d^4 x\, \left[\sqrt{-\hat{g}} e^{-\frac{2}{3}\sigma}\right]\,
\left\{\left[\frac{\hat{R}}{2\kappa} + \hat{{\cal L}}\right]\,e^{\frac{2}{3}\sigma}\right\},
\end{equation}
and if we require that $\delta {\cal S}_{EH} =0$, we obtain
\begin{equation}
-\frac{\delta V}{V} = \frac{\delta \left[\frac{\hat{R}}{2\kappa} + \hat{{\cal L}}\right]}{\left[\frac{\hat{R}}{2\kappa} + \hat{{\cal L}}\right]}
= \frac{2}{3} \,\delta\sigma,
\end{equation}
where $\delta\sigma = \sigma_{\mu} dx^{\mu}$ is an exact differential and $V=\sqrt{-\hat{ g}}$ is the volume of the Riemannian manifold.
The relativistic quantum algebra is given by\cite{rb,rb1}
\begin{equation}\label{con}
\left[\sigma(x),\sigma^{\alpha}(y) \right] =- i \Theta^{\alpha}\, \delta^{(4)} (x-y), \qquad \left[\sigma(x),\sigma_{\alpha}(y) \right] =
i \Theta_{\alpha}\, \delta^{(4)} (x-y),
\end{equation}
with $\Theta^{\alpha} = i \hbar\, \hat{U}^{\alpha}$ and $\Theta^2 = \Theta_{\alpha}
\Theta^{\alpha} = \hbar^2 \hat{U}_{\alpha}\, \hat{U}^{\alpha}$ for the Riemannian components of velocities $\hat{U}^{\alpha}$. Finally, the metric tensor (in cartesian coordinates), with back-reaction effects included holds (here $\bar{g}_{\alpha\beta}$ are the components of the background metric tensor):
\begin{equation}\label{met1}
g_{\mu\nu} = {\rm diag}\left[\bar{g}_{00}\, e^{2\sigma/3}, \bar{g}_{11}\, e^{-2\sigma/3}, \bar{g}_{22}\, e^{-2\sigma/3},\bar{g}_{33}\, e^{-2\sigma/3}\right],
\end{equation}
which preserves the invariance of the E-H action.

\section{Back-reaction solution in a de Sitter metric}\label{3}

We consider a static de Sitter line element written in spherical coordinates
\begin{equation}\label{m}
ds^2 = f(r) dt^2 - \frac{1}{f(r)} dr^2 - r^2 d\Omega^2,
\end{equation}
where $d\Omega^2 = d\theta^2 + \sin^2(\theta) \,d\phi^2$ and $f(r) = 1- (r/\alpha)^2$, such that $H$ is the Hubble parameter, $c$ is the light velocity in the vacuum, and $\alpha= c/H$ is the Hubble horizon, which is related to the cosmological constant $\Lambda = 3 (H/c)^2$. In order for describe the back-reaction effects in the interior of the de Sitter space, we must consider solutions of the equation $\Box\sigma=0$, for $r<\alpha$, where the space-time is 4D hyperbolic with signature $(+,-,-,-)$, due to the fact $f(r)>0$. The massless scalar field $\sigma$ for the line element \eqref{m} and $r<\alpha$ is described by the equation
\begin{equation}
\frac{1}{f(r)} \frac{\partial^2 \sigma}{\partial t^2}+\frac{1}{r^{2}} \frac{\partial}{\partial r} \left[r^{2}f(r)\frac{\partial \sigma}{\partial r}\right]+\frac{1}{r^{2}\,\sin(\theta)}\frac{\partial}{\partial\theta} \left[\sin(\theta) \frac{\partial \sigma}{\partial\theta}\right]+\frac{1}{{\sin^2(\theta)}}\frac{\partial^2 \sigma}{\partial\phi^2} =0.
\end{equation}
Because there are a finite number of states that describe the interior of a de Sitter space-time, we can expand the field $\sigma$ as a superposition $\sigma_{(n,l,m)}(t,r,\theta,\phi) \sim R_{(n,l)}(r)\,{\tau}_{n}(t)\,Y_{(l,m)}(\theta,\phi)$, where the functions $Y_{(l,m)}(\theta,\phi)$ are the usual spherical harmonics. In this case, the radial equation for $R_{(n,l)}(r)$ and temporal one for ${\tau}_{n}(t)$, are given by
\begin{eqnarray}
\frac{\partial^{2}{\tau}_{n}}{\partial t^{2}}&+&\left(\frac{E_{(n,l)}}{\hbar}\right)^2\,{\tau}_n(t)=0, \label{Temporal} \\
r^{2}f(r)^2\frac{\partial^{2}R_{(n,l)}}{\partial r^{2}}&=&\left[l(l+1)f(r)+\left(\frac{E_{(n,l)}}{\hbar}\right)^2\,r^2\right]\,R_{(n,l)}-\left(r^2f(r)\frac{df}{dr}+2r\,f(r)^2\right)\frac{\partial R_{(n,l)}}{\partial r}. \nonumber \\ \label{Radial}
\end{eqnarray}
We shall use the variable substitution $r=\frac{\alpha}{c}\sqrt{u}$, that implies that $0\leq u<1$. With this replacement we obtain that the solution of the radial equation can be expressed in terms of the confluent hypergeometric functions: $_2F_1\left([a,b],[c],u\right)$
\begin{equation}\label{RadialSolution}
R_{(n,l)}(u)=C_1(u-1)^{\frac{\alpha E_{(n,l)}}{2 \hbar c}}u^{l/2}\,_2F_1\left([a_1,b_1],[c_1],u\right)+C_2(u-1)^{\frac{\alpha E_{(n,l)}}{2 \hbar c}}u^{-\frac{l+1}{2}}\,_2F_1\left([a_2,b_2],[c_2],u\right),
\end{equation}
where $0<u<1$ and the parameters $a_1$, $b_1$, $c_1$, $a_2$, $b_2$ and $c_2$ are given by
\begin{equation}
a_1=\frac{\alpha E_{(n,l)}}{2 \hbar c}+\frac{l}{2}, \quad b_1=a_1+\frac{3}{2}, \quad c_1=\frac{3}{2}+l, \quad a_2=\frac{\alpha E_{(n,l)}}{2 \hbar c}-\frac{l+1}{2}, \quad b_2=a_2+\frac{3}{2}, \quad c_2=\frac{1}{2}-l.
\end{equation}

The series representation of $_2F_1\left([a,b],[c],u\right)$ determines if either $a$ or $b$ is a non-positive integer $-n$, in which case the function is reduced to a polynomial of order $n$. Our aim is using this condition in order for motivate validity of the uncertainly principle for each energetic level and the discretization of the $\alpha$ values, in order for relate the solutions (\ref{RadialSolution}) to the recently studied Schwarzschild black hole's mass case\cite{mba}.

\subsection{Uncertainly principle, Energy levels and the cosmological constant}

In order for avoiding divergent solutions of $_2F_1\left([a_1,b_1],[c_1],u\right)$ with $\alpha_{nl}<0$, we shall propose $C_1=0$ in (\ref{RadialSolution}). Furthermore the condition $a_2=-n$ in (\ref{RadialSolution}), gives us
\begin{equation}\label{elevels}
\frac{E_{(n,l)}}{2}\frac{\alpha_{(n,l)}}{c}=\frac{\hbar}{2},
\end{equation}
with $\alpha_{(n,l)}=\frac{\alpha}{l+1-2n}$.

The expression (\ref{elevels}) is very important because it tells us that the uncertainly principle is fulfilled for each energy level. We can see that for this expression that the admissible energy-levels are inversely proportional to the cosmological horizon $c/H$. Furthermore, this also provides a discretization of the cosmological constant $\Lambda=\frac{3}{\alpha^2}$: $\Lambda_{(n,l)}$, in terms of the eigenvalues $n$ and $l$. Therefore, if we introduce $\Lambda=\frac{3}{\alpha^2}$ and
\begin{equation}\label{la}
\Lambda_{(n,l)}=\frac{3}{\alpha_{(n,l)}^2},
\end{equation}
we obtain a relation between $\Lambda_{(n,l)}$ and the energy levels $E_{(n,l)}$:
\begin{equation}
\Lambda_{(n,l)}=\frac{3E_{(n,l)}^2}{c^2\hbar^{2}}.
\end{equation}
With these results, now the parameters of the radial solution $R_{(n,l)}(u)$ can be written as
\begin{equation}
R_{(n,l)}(u)=C(u-1)^{\frac{l}{2}-(n+1)}u^{-\frac{l+1}{2}}\,_2F_1\left([-n,-n+\frac{3}{2}],[\frac{1}{2}-l],u\right),
\end{equation}
where $C$ is a constant to be determined by normalization.

We shall suppose that exists a lower bound for the energy corresponding to the Planck energy: $E_p=m_p\,c^2$, where $m_p$ is the Planck mass. Hence, taking into account all the possible values $E_{(n,l)}$, and using the expression (\ref{elevels}), we obtain the condition for the allowed $l$-values:
\begin{equation}
2n+\frac{\alpha\,m_{p}\,c^2}{\hbar}-1=2n+N(\alpha)-1\leq\,l,
\end{equation}
where $N(\alpha)=\frac{\alpha\,m_{p}\,c^2}{\hbar}\geq1$, once we consider the Planck length: $l_p$, and we assume $\alpha_{(n,l)}\,\geq\,l_{p}$. The limit case for the previous expression corresponds to $N(\alpha)=1$ and $l=2n$, and it's consistent with the definition for $\alpha_{(n,l)}$ in (\ref{elevels}).

Finally, we can write the complete solution for the field $\sigma(t,r,\theta,\phi)$, as
\begin{equation}
\sigma(t,r,\theta,\phi)= \sum_{n=0}^{N-1}\,\sigma_n(t,r,\theta,\phi),
\end{equation}
where
\begin{eqnarray}
\sigma_n(t,r,\theta,\phi)= \sum_{l\geq L_{-}}^{L_{+}}\,\sum_{m =-l}^{l} \left[A_{(n,l,m)}\,\sigma_{(n,l,m)}(t,r,\theta,\phi)+ A^{\dagger}_{(n,l,m)}\,\sigma^*_{(n,l,m)}(t,r,\theta,\phi)\right],
\end{eqnarray}
such that the modes $\sigma_{(n,l,m)}(t,r,\theta,\phi)$, are:
\begin{equation}
\sigma_{(n,l,m)}(t,r,\theta,\phi) = \left(\frac{E_{(n,l)}}{\hbar}\right)^2 \, R_{(n,l)}(r)\, Y_{(l,m)}(\theta,\phi)\, {\tau}_n(t),
\end{equation}
with a radial local solution expanded around $u=0$.

\subsection{de Sitter Temperature from RQG}

In order for calculating the temperature inside the horizon, we shall consider a discrete transition from $\alpha_{(0,0)}$ to $\alpha_{(n,l)}$: $\Delta \alpha_{(n,l)}=\alpha_{(n,l)}-\alpha_{(0,0)}$, and another one for the entropy $\Delta{S_{(n,l)}}=S_{(n,l)}-S_{(0,0)}$. In this framework, we shall define the level dependent temperature
\begin{equation}
T_{(n,l)}=\frac{\Delta \alpha_{(n,l)}}{\Delta S_{(n,l)}} \left(\frac{K_B}{\hbar\,c}\right),
\end{equation}
where $S_{(n,l)}=\frac{A_{(n,l)} }{4}\,\left(\frac{K_B}{\hbar\,c}\right)^2$ and the area related to horizon of each level: $\alpha_{(n,l)}$, will be
\begin{equation}
A_{(n,l)}=4\pi\alpha_{(n,l)}^2,
\end{equation}
which can be related with $\Lambda_{(n,l)}$ using (\ref{la}):  $A_{(n,l)}={12\pi \over \Lambda_{(n,l)}}$. Therefore, the entropy for each $(n,l)$-values, is
\begin{equation}
S_{(n,l)} = \frac{3\pi}{\Lambda_{(n,l)}} \left(\frac{K_B}{\hbar\,c}\right)^2= \pi\,\alpha_{(n,l)}^2 \left(\frac{K_B}{\hbar\,c}\right)^2 =\frac{\sqrt{3} \pi }{E^2_{(n,l)}} {K_B}^2,
\end{equation}
which is inversely proportional to the squared value of the $(n,l)$-energy. After specializing, we obtain that the value of a generic $T_{(n,l)}$ corresponds to
\begin{equation}
T_{(n,l)}=\frac{l+1-2n}{\alpha\pi(l-2n+2)}\left(\frac{\hbar\,c}{K_B}\right).
\end{equation}

If we suppose that $E_{(n,l)}>0$ and $T_{(n,l)}>0$, the possible values of $l$, corresponding to each $n$-value, will be \begin{equation}\label{cond}
l(n)>l_{min}(n)=2(n-1).
\end{equation}
Therefore, from the condition (\ref{cond}) we obtain that for a given $n$, $l$ must take the values $l \geq 2n$.
Furthermore, all combinations $l=2n$ guarantee $T_{(n,l=2n)}=T_{BH}$, where $T_{BH}=\frac{\hbar\,c}{2\pi\alpha\,K_B}$ is the Bekenstein-Hawking (B-H) temperature\cite{bek,hawking} and $K_B$ is the Boltzmann constant. In particular, $T_{(0,0)}=T_{(1,2)} = ...= T_{BH}$.

This means that we can study the interior levels by choosing for each ($l=2n$)-level, given by $l=2n+m$, with $m(l)\geq 0$, the temperature:
\begin{equation}
T_{(n,2n+m)}=\frac{m+1}{(m+2)\pi\alpha} \left(\frac{\hbar\,c}{K_B}\right)\equiv T_m
\end{equation}
For $m=0$, we have the exterior level $T_{m=0}=T_{BH}$, and for large values $l$ we have $T_{m\rightarrow\infty}=2\,T_{BH}$. This is the same behavior that energy levels in the Schwarzschild's Black Hole interior\cite{mba}. The important here is that this behavior is repeated for all the
possible values of $m(l)$, on each ($l=2n$)-level.

Finally, we can consider the difference of temperature between two consecutive levels of each ($l=2n$)-level: $\Delta T_m = T_{m+1}-T_{m}$.
We obtain
\begin{equation}
\Delta T_{m} = \frac{1}{\pi\alpha\,(m+2)\,(m+3)}\left(\frac{\hbar\,c}{K_B}\right),
\end{equation}
such that the summation on all the possible values of $\Delta T_{m}$, on each ($l=2n$)-level results to be the BH-temperature:
\begin{equation}
\lim_{L\rightarrow\infty} \sum\limits_{m=0}^{L} \Delta T_m = \frac{1}{\pi\alpha}\left(\frac{\hbar\,c}{K_B}\right) \lim_{L\rightarrow\infty} \left(\frac{1}{2}-\frac{1}{L+3}\right)=T_{BH},
\end{equation}
where $L$ is the maximum value of $m$. This is a very important result that replies whole obtained in \cite{mba}, but here on each ($l=2n$)-level. Notice that the extreme case where $\alpha=l_p$ is the Planck length, is a good candidate to describe the initial state of the universe in a de Sitter metric, with a B-H temperature:
\begin{equation}
T_{BH} = \frac{\hbar \,c}{2\pi K_B\,l_p},
\end{equation}
and a Planck energy: $E_p= m_p\,c^2$.

In a general case, if we suppose that the universe is described by a de Sitter expansion, which is believed will occur in the late state of the evolution of the universe, one could calculate the minimum $T_{BH}$ (which cannot be understood as the temperature of the Background Cosmic Radiation) in the universe. Using the fact that $\Lambda=3/\alpha^2$, and $\alpha=c/H$, we could write the background metric in terms of the B-H temperature:
\begin{equation}
f(r)= 1- \left(\frac{r}{\alpha}\right)^2 = 1 - \left(\frac{2\pi\,K_B\,T_{BH}}{\hbar \,c}\,r\right)^2,
\end{equation}
with $T_{BH} = {H \hbar \over 2 \pi \,K_B}$. Notice that $H$ is a cosmological observable, so that one could immediately calculate the B-H temperature in the universe.

\section{Final comments}\label{4}

In this work we have studied back-reaction effects in the interior of a de Sitter space-time ({\it i. e.}, for $r<c/H$), using the RQG formalism in which we take into account, when we variate the EH action, the flux that cross the 3D-gaussian hypersurface. The extended manifold is obtained by making a displacement from the background Riemann manifold to the new extended manifold (\ref{ConexionWeyl}). This flux is described by a scalar field $\sigma$ (more precisely by their partial derivatives: $\sigma^{\alpha}$), that describes back-reaction effects of the space-time, so that the metric tensor with back-reaction effects included are given by (\ref{met1}). We have applied this formalism to study the back-reaction effects on sub-horizon scales of a static de Sitter metric, and, for the radial solution, we have found a finite number of discrete energy levels for the $l=2n$ values, such that, for each energy level, we have a number $L$ of possible values $L \geq  l\geq 2n $, for a scalar field solution, obtained from a polynomial condition of the confluent hypergeometric functions, which is expanded around $u=0$. The interesting is that we recover the same structure for the temperature values, that in the interior of the Schwarzschild black-hole\cite{mba}, but here for each ($l=2n$)-level. Furthermore, the uncertainty principle (\ref{elevels}), is valid for each energy level on sub-horizon scales of the space-time, and the temperature and entropy are dependent on the number of sub-states with different $m(l)$, on such scales. When this number tends to infinity: $L \rightarrow \infty$, we recover the B-H temperature for this ($l=2n$)-level: $\lim_{L\rightarrow\infty} \sum\limits_{m=0}^{L} \Delta T_m =T_{BH}$. Therefore, we propose that the primordial universe could be described by a Planck energy and a B-H temperature: $T_{BH}=\left(\frac{\hbar\,c}{2\pi\,l_p\,K_B}\right)$ in a de Sitter space-time\cite{muk1,muk2,vil}. Our approach could be extended to another in which we describe a evolving geometry such that $\Lambda$ is a time-decaying cosmological parameter on the Riemann manifold. This issue, which is beyond the scope of this work, will be developed in the future.

\section*{Acknowledgements}

\noindent The authors acknowledge CONICET, Argentina (PIP 11220150100072CO) and UNMdP (EXA852/18) for financial support.


\end{document}